\documentstyle[a4,12pt]{article}
\renewcommand{\baselinestretch}{1.25}

\begin{document}

\begin{flushright}
{\bf Preprint LMU-98-11} \\
hep-ph/9810249 \\
{October 1998}
\end{flushright}

\vspace{0.2cm}

\begin{center}
{\large\bf Opposite-sign Dilepton Asymmetry of Neutral $B$ Decays: \\
a Probe of New Physics from $CPT$ or $\Delta B = \Delta Q$ Violation}
\end{center}

\vspace{1.2cm}

\begin{center}
{\bf Zhi-zhong Xing}\footnote{
E-mail: Xing$@$hep.physik.uni-muenchen.de }\\
{\sl Sektion Physik, Universit${\sl\ddot a}$t M${\sl\ddot
u}$nchen, Theresienstrasse 37A, 80333 M${\sl\ddot u}$nchen, Germany}
\end{center}

\vspace{2cm}

\begin{abstract}
We show that new physics, either from $CPT$ violation in 
$B^0$-$\bar{B}^0$ mixing or from $\Delta B = -\Delta Q$
transitions, may lead to an opposite-sign dilepton
asymmetry of neutral $B$-meson decays. Both effects
have the same time-dependent behavior and therefore
are in general indistinguishable from each other. We also clarify
some ambiguity associated with the constraint on $CPT$ violation 
in semileptonic decays of incoherent $B^0$ and $\bar{B}^0$ mesons.
\end{abstract}

%\vspace{2cm}
%\begin{center}
%PACS number(s): 11.30.Er, 13.25.Hw, 14.40.Nd
%\end{center}

\newpage

The $B$-meson factories under construction are going to provide
a unique opportunity for the study of $CP$ violation, both
its phenomenon and its origin. At the second or
final stage of these 
facilities sufficient $B\bar{B}$ events (e.g.,
${\cal N}_{B\bar{B}} \geq 10^9$) will be accumulated,
then a direct test of the $CPT$ symmetry and the $\Delta B
= \Delta Q$ rule should be experimentally feasible. 
This is expected to open a new
window for probing new physics beyond the standard model, 
as both $CPT$ and $\Delta B = \Delta Q$ conservation laws 
work extremely well within the standard model. 
Up to now some phenomenological analyses 
of possible $CPT$-violating signals in semileptonic
and nonleptonic decays of neutral $B$ mesons have been made
in Refs. \cite{Sanda} -- \cite{Sakai98}, whose results depend
on the validity of the $\Delta B = \Delta Q$ rule.
The possibility to detect the effect of $\Delta B =-\Delta Q$
transitions at $B$ factories has been 
investigated in Ref. \cite{Sarma} in the assumption of
$CPT$ invariance.

It was noticed in Refs. \cite{Sanda} -- \cite{Sakai98}
that $CPT$ violation in
$B^0$-$\bar{B}^0$ mixing could lead to an opposite-sign
dilepton asymmetry of neutral-$B$ decays, or more
generally, an
asymmetry between 
$B^0 (t)\rightarrow B^0 \rightarrow l^+X^-$ and
$\bar{B}^0 (t)\rightarrow \bar{B}^0 \rightarrow l^-X^+$ decay rates.
Applying this idea to the time-dependent semileptonic
decays of $B^0$ and $\bar{B}^0$ mesons at the
$Z$ resonance, the OPAL Collaboration obtained a
constraint on the $CPT$-violating parameter $\delta_B$, i.e.,
${\rm Im} \delta_B =-0.020 \pm 0.016 \pm 0.006$ \cite{OPAL}.
The ideal experimental environment for measuring the
time distribution of opposite-sign dilepton events will
be at the KEK and SLAC asymmetric $B$-meson factories, where
$B^0 \bar{B}^0$ pairs can coherently be produced
through the decay of the $\Upsilon (4S)$ resonance.

Allowing both $CPT$ violation and $\Delta B = -\Delta Q$
transitions, we shall carry out a new analysis of the
opposite-sign dilepton asymmetry of neutral $B$ decays
and clarify some ambiguity
associated with the constraint on $CPT$ violation in semileptonic
decays of incoherent $B^0$ and $\bar{B}^0$ mesons.
It is shown that the effect of $\Delta B = -\Delta Q$ transitions
and that of $CPT$ violation have the same time-dependent
behavior in the opposite-sign dilepton events,
therefore they are in general indistinguishable from each other.
We point out that a meaningful constraint on
${\rm Im}\delta_B$ depends actually on the smallness
of the decay width difference between two neutral-$B$ 
mass eigenstates, on the condition 
$|{\rm Re}\delta_B| \leq |{\rm Im}\delta_B|$, and
on the validity of the $\Delta B = \Delta Q$ rule.

\vspace{0.5cm}

Mixing between $B^0$ and $\bar{B}^0$ mesons arises naturally
from their common coupling to a subset of real and
virtual intermediate states, hence the mass eigenstates
$|B_1\rangle$ and $|B_2\rangle$ are different from the
flavor eigenstates $|B^0\rangle$ and $|\bar{B}^0\rangle$.
A {\it non-linear} parametrization of $CP$- and $CPT$-violating
effects in $B^0$-$\bar{B}^0$ mixing reads as
\begin{eqnarray}
|B_1\rangle & = & \cos\frac{\theta}{2} ~ e^{-{\rm i}\frac{\phi}{2}}
|B^0\rangle ~ + ~ \sin\frac{\theta}{2} ~ e^{+{\rm i}\frac{\phi}{2}}
|\bar{B}^0\rangle \; , \nonumber \\
|B_2\rangle & = & \sin\frac{\theta}{2} ~ e^{-{\rm i}\frac{\phi}{2}}
|B^0\rangle ~ - ~ \cos\frac{\theta}{2} ~ e^{+{\rm i}\frac{\phi}{2}}
|\bar{B}^0\rangle \; , 
%		(1)
\end{eqnarray}
where $\theta$ and $\phi$ are in general complex, and the
normalization factors of $|B_1\rangle$ and $|B_2\rangle$ 
have been neglected. $CPT$ invariance requires $\cos\theta =0$,
while $CP$ conservation requires both $\cos\theta =0$ and
$\phi =0$ \cite{Lee}. In some literature the {\it linear} parametrization
%%%%%%%%%%%%%%%%%%%
\footnote{Here we use the notation
$(\delta_B, \epsilon^{~}_B)$ instead of
$(\delta, \epsilon)$, as the latter is commonly adopted
to describe $CPT$- and $CP$-violating effects in the
$K^0$-$\bar{K}^0$ mixing system.}
%%%%%%%%%%%%%%%%%%%%
\begin{eqnarray}
|B_1\rangle & = & (1 + \epsilon^{~}_B + \delta_B ) |B^0\rangle ~ 
+ ~ (1-\epsilon^{~}_B -\delta_B) |\bar{B}^0\rangle \; , \nonumber \\
|B_2\rangle & = & (1+\epsilon^{~}_B -\delta_B) |B^0\rangle ~ 
- ~ (1-\epsilon^{~}_B +\delta_B) |\bar{B}^0\rangle \; , 
%		(2)
\end{eqnarray}
together with the conventions $|\epsilon^{~}_B|\ll 1$ and
$|\delta_B|\ll 1$, has been adopted (here again the normalization
factors of $|B_1\rangle$ and $|B_2\rangle$ are neglected).
It is straightforward to find the relationship between
$(\theta,\phi)$ and $(\delta_B,\epsilon^{~}_B)$ parameters in the
leading-order approximation, $\epsilon^{~}_B \approx {\rm i} \phi/2$
and $\delta_B \approx \cot\theta/2$, provided the conventions
$|\cos\theta|\ll 1$ and $|\phi|\ll 1$ are taken. 
Note that the requirement $|\epsilon^{~}_B|\ll 1$ (or $|\phi|\ll 1$)
corresponds to an exotic phase convention of quark fields, which
is incompatible with several popular parametrizations of 
the quark flavor mixing matrix \cite{Nir}. In subsequent calculations
we shall mainly make use of the $(\theta, \phi)$ notation 
(in which $\phi$ may take arbitrary values), and translate it
into the $(\delta_B, \epsilon^{~}_B)$ notation when necessary.

The proper-time evolution of an initially pure
$|B^0\rangle$ or $|\bar{B}^0\rangle$ state is then given 
as \cite{Xing94}
\begin{eqnarray}
|B^0(t)\rangle & = & e^{- \left ({\rm i} m +\frac{\Gamma}{2}
\right ) t} \left [g^{~}_+(t) |B^0\rangle ~ + ~
\tilde{g}^{~}_+(t) |\bar{B}^0\rangle \right ] \; ,
\nonumber \\
|\bar{B}^0(t)\rangle & = & e^{- \left ({\rm i} m +\frac{\Gamma}{2}
\right ) t} \left [g^{~}_-(t) |\bar{B}^0\rangle ~ + ~
\tilde{g}_-(t) |B^0\rangle \right ] \; ,
%		(3)
\end{eqnarray}
where 
\begin{eqnarray}
g^{~}_{\pm}(t) & = & \cosh \left (\frac{{\rm i}x -y}{2}
\Gamma t \right ) ~ \pm ~ \cos\theta
\sinh \left (\frac{{\rm i}x -y}{2}
\Gamma t \right ) \; , \nonumber \\
\tilde{g}^{~}_{\pm}(t) & = & \sin\theta ~ e^{\pm {\rm i}\phi}
\sinh \left (\frac{{\rm i} x - y}{2} \Gamma t \right ) \; .
%		(4)
\end{eqnarray}
In Eqs. (3) and (4) we have defined $m\equiv (m_1 + m_2)/2$,
$\Gamma \equiv (\Gamma_1 + \Gamma_2)/2$, $x \equiv (m_2-m_1)/
\Gamma$ and $y\equiv (\Gamma_1 -\Gamma_2)/(2\Gamma)$, where 
$m_{1,2}$ and $\Gamma_{1,2}$ are the mass and width of 
$B_{1,2}$. The mixing parameter $x
\approx 0.7$ has been measured \cite{PDG98}, and $y \sim O(10^{-2})$
is theoretically expected \cite{Bigi}.
To calculate the time distribution of opposite-sign dilepton
events on the $\Upsilon (4S)$ resonance, 
we neglect possible tiny effects from
electromagnetic final-state interactions and assume $CPT$
invariance in the direct transition amplitudes of semileptonic
$B$ decays. 
Such an assumption
can be examined, without the mixing-induced complexity, 
by measuring the charge
asymmetry of semileptonic $B^{\pm}$ decays.
The effect of possible
$\Delta B =-\Delta Q$ transitions need be taken into account.
We then write the relevant semileptonic decay amplitudes
as follows:
\begin{eqnarray}
\langle l^+|B^0\rangle & = & A_l \; ,
~~~~~~~~
\langle l^+|\bar{B}^0\rangle \; =\; \sigma^{~}_l
A_l \; ; \nonumber \\
\langle l^-|\bar{B}^0\rangle & = & A^*_l \; ,
~~~~~~~~
\langle l^-|B^0\rangle \; =\; \sigma^*_l
A^*_l \; ,
%		(5)
\end{eqnarray}
where $\sigma^{~}_l$ measues the $\Delta B =-\Delta Q$ 
effect and $|\sigma^{~}_l|\ll 1$ holds.

$|\sigma^{~}_l| \neq 0$ implies that it is in practice impossible to
have a {\it pure} tagging of the $B^0$ or $\bar{B}^0$ state through 
its semileptonic decay (to $l^+$ or $l^-$). For $B^0\bar{B}^0$
pairs produced {\it incoherently} on the $Z$ resonance or elsewhere,
the assumption $|\sigma^{~}_l| =0$ (i.e., the exact $\Delta B = \Delta Q$
rule) is indeed a necessary condition to get pure flavor tagging of $B^0$
and $\bar{B}^0$ mesons and to study possible tiny violation of
$CPT$ symmetry in their semileptonic decays
%%%%%%%%%%%%%%%
\footnote{Note that the treatment in Ref. \cite{Kos95}, which
takes $|\sigma^{~}_l| \neq 0$ on the one hand but assumes the pure
flavor tagging of $B^0$ and $\bar{B}^0$ states on the other hand,
is controversial in the experimental environment where $B^0\bar{B}^0$
pairs are incoherently produced.}.
%%%%%%%%%%%%%%%%%

The pure flavor tagging is however unnecessary 
for the study of fine effects induced by $CPT$ violation and (or)
$\Delta B = -\Delta Q$ transitions in opposite-sign dilepton products
of {\it coherent} $B^0\bar{B}^0$ pairs on the $\Upsilon (4S)$ 
resonance. 
Here the decay of one neutral $B$ meson at proper time $t_1$
into a semileptonic state 
(e.g., $e^{\pm}X^{\mp}_e$) may only serve as a rough flavor tagging
for the other meson decaying at proper time $t_2$
into another semileptonic
state (e.g., $\mu^{\mp} X^{\pm}_\mu$). The overall final state
is therefore an opposite-sign dilepton event. Since 
we are only interested in how the decay rate depends on
the time difference $t_2 -t_1$ at an asymmetric
$B$ factory, we just take $t_1 =0$ and
$t_2 =t$ with the convention $t>0$ for 
the wave function of the coherent $B^0\bar{B}^0$ pair:
\begin{equation}
\Psi (t) \; =\; \frac{1}{\sqrt{2}} \left [
|B^0(0) \rangle \otimes |\bar{B}^0(t)\rangle - 
|B^0(t)\rangle \otimes |\bar{B}^0(0)\rangle \right ] \; . 
%		(6)
\end{equation}
Then the rates of two opposite-sign dilepton decays
(with $l^{\pm}$ events at $t=0$) read
\begin{eqnarray}
{\cal R}(t) & \equiv & |\langle l^+l^{'-}|\Psi (t)\rangle|^2
\; =\; \frac{1}{2} |A_l|^2 |A_{l'}|^2 e^{-\Gamma t} 
\left |g^{~}_-(t) + \tilde{g}^{~}_-(t)\sigma^*_{l'} -
\tilde{g}^{~}_+(t) \sigma^{~}_l \right |^2 \; , \nonumber \\
\bar{\cal R}(t) & \equiv & |\langle l^-l^{'+}|\Psi (t)\rangle|^2
\; =\; \frac{1}{2} |A_l|^2 |A_{l'}|^2 e^{-\Gamma t} 
\left |g^{~}_+(t) + \tilde{g}^{~}_+(t)\sigma^{~}_{l'} -
\tilde{g}^{~}_-(t) \sigma^*_l \right |^2 \; ,
%		(7)
\end{eqnarray}
where only the contributions of $O(|\sigma^{~}_l|)$
and $O(|\sigma^{~}_{l'}|)$ are kept, and
the functions $g^{~}_{\pm}(t)$ and $\tilde{g}^{~}_{\pm}(t)$ can be
found in Eq. (3). We arrive finally at
\begin{eqnarray}
{\cal R}(t) & = & \frac{1}{4} |A_l|^2 |A_{l'}|^2 e^{-\Gamma t}
\left [\cosh (y\Gamma t) 
~ + ~ 2{\rm Re}\Omega \sinh (y\Gamma t) \right . \nonumber \\
&  & ~~~~~~~~~~~~~~~~~~~ \left . + ~ \cos (x\Gamma t) 
~ + ~ 2{\rm Im}\Omega \sin (x\Gamma t) \right ] \; ,
\nonumber \\
\bar{\cal R}(t) & = & \frac{1}{4} |A_l|^2 |A_{l'}|^2 e^{-\Gamma t}
\left [\cosh (y\Gamma t) 
~ - ~ 2{\rm Re}\bar{\Omega} \sinh (y\Gamma t) \right . \nonumber \\
&  & ~~~~~~~~~~~~~~~~~~~ \left . + ~ \cos (x\Gamma t)  
~ - ~ 2{\rm Im}\bar{\Omega} \sin (x\Gamma t) \right ] \; , 
%		(8)
\end{eqnarray}
where
\begin{eqnarray}
\Omega & = & \cos\theta ~ + ~ \xi_{ll'} \sin\theta \; , \nonumber \\
\bar{\Omega} & = & \cos\theta ~ + ~ \bar{\xi}_{ll'} \sin\theta \; ,
%		(9)
\end{eqnarray}
and
\begin{eqnarray}
\xi_{ll'} & = & \sigma^{~}_l e^{+{\rm i}\phi} ~ - ~
\sigma^*_{l'} e^{-{\rm i}\phi} \; , \nonumber \\
\bar{\xi}_{ll'} & = & \sigma^{~}_{l'} e^{+{\rm i}\phi} ~ - ~
\sigma^*_l e^{-{\rm i}\phi} \; .
%		(10)
\end{eqnarray}
In obtaining Eq. (8) we have neglected the contributions
of $O(|\Omega|^2)$ and $O(|\bar{\Omega}|^2)$. 

Obviously the $\Delta B =-\Delta Q$ effects are
signified by the rephasing-invariant
parameters $\xi_{ll'}$ and $\bar{\xi}_{ll'}$.
$\xi_{ll'} = \bar{\xi}_{ll'}$ holds for $l'=l$, and
$\xi_{ll'} = -\bar{\xi}^*_{ll'}$ holds if ${\rm Im}\phi =0$.
We find that $\xi_{ll'}$ or $\bar{\xi}_{ll'}$ has the
same time-dependent behavior as the $CPT$-violating 
parameter $\cos\theta$ in the decay rate ${\cal R}(t)$ or
$\bar{\cal R}(t)$. {\it This important feature implies that it is 
in general impossible to distinguish between the effects of
$CPT$ violation and $\Delta B =-\Delta Q$ transitions in
the opposite-sign dilepton events, unless one of them 
is remarkably smaller than the other}. 

\vspace{0.5cm}

For illustration we simplify the result in Eq. (8) by
taking $\sigma^{~}_l = \sigma^{~}_{l'} =0$, i.e,
no $\Delta B = -\Delta Q$ transition involved.
In this case one gets
\begin{eqnarray}
{\rm Re}(\Omega + \bar{\Omega}) & = & 2 {\rm Re}(\cos\theta) \; ,
\nonumber \\
{\rm Im}(\Omega + \bar{\Omega}) & = & 2 {\rm Im}(\cos\theta) \; ,
%		(11)
\end{eqnarray}
which results in an asymmetry between
the decay rates ${\cal R}(t)$ and $\bar{\cal R}(t)$.
Defining ${\cal A}(t)$ as the ratio of 
${\cal R}(t) - \bar{\cal R}(t)$ to
${\cal R}(t) + \bar{\cal R}(t)$, we then have
\begin{equation} 
{\cal A}(t) \; = \; 2\frac{{\rm Re}(\cos\theta) \sinh (y\Gamma t) 
+ {\rm Im}(\cos\theta) \sin (x\Gamma t)}
{\cosh (y\Gamma t) + \cos (x\Gamma t)} \; \; ,
%		(12)
\end{equation}
which is independent of the $CP$-violating
parameter $\phi$. This formula has already been obtained in
Refs. \cite{Xing97} and \cite{Sakai98}
%%%%%%%%%%%%%%%%%%%%%%%
\footnote{Note that the definitions of $y$ in Refs. \cite{Xing97}
and \cite{Sakai98} are different in sign. Taking 
this into account, one can get full consistency between
the results in these two papers.}.
%%%%%%%%%%%%%%%%%%%%%%
It should be noted that the asymmetry ${\cal A}(t)$, 
due to its time dependence,
is not restricted to the range $[-1, +1]$. In the
limit $y=0$, Eq. (12) turns out to be
${\cal A}(t) = 2{\rm Im}(\cos\theta) \tan (x\Gamma t/2)$,
which becomes infinity on the point $\Gamma t=\pi/x$
(i.e., around $\Gamma t \approx 4.5$).

If one translates $\cos\theta$ into $\delta_B$, then
${\rm Re}(\cos\theta) \approx 2{\rm Re}\delta_B$ and
${\rm Im}(\cos\theta)\approx 2{\rm Im}\delta_B$. Therefore
a constraint on ${\rm Im}\delta_B$ is achievable from
Eq. (12) with $y=0$. This is indeed the case taken by
the OPAL Collaboration in their measurement \cite{OPAL}, where
the rate difference between $B^0(t)\rightarrow
B^0\rightarrow l^+X^-$ and $\bar{B}^0(t)\rightarrow \bar{B}^0
\rightarrow l^-X^+$
transitions is essentially 
equivalent to the asymmetry of opposite-sign
dilepton events of $B^0\bar{B}^0$ decays under
discussion. As the theoretical calculation favors
$y/x \sim 10^{-2}$ \cite{Bigi}, the approximation 
$y \approx 0$
made in Ref. \cite{OPAL} seems quit reasonable.

However, such an approximation may not be valid if
$|{\rm Re}\delta_B| \gg |{\rm Im}\delta_B|$ holds. 
Although there is no experimental information about the
relative magnitude of ${\rm Re}\delta_B$ and ${\rm Im}\delta_B$,
a string theory with spontaneous $CPT$ violation and
Lorentz symmetry breaking predicts \cite{Kos95}
\begin{equation}
r \; \equiv \; \frac{{\rm Re}\delta_B}{{\rm Im}\delta_B} \; =\; 
\pm \frac{x}{y} \; \sim \; \pm 10^{2} \; .
%		(13)
\end{equation}
In this case, the $\sinh (y\Gamma t)$
and $\sin (x\Gamma t)$ terms in the numerator of
${\cal A}(t)$ are comparable in magnitude even for
small $t$. Then a meaningful constraint on
${\rm Im}\delta_B$ becomes impossible. To illustrate,
we plot the time distribution of ${\cal A}(t)$
in Fig. 1 with the inputs 
$r = \pm 1$ and $r=\pm 100$ (for fixed ${\rm Im}\delta_B$), 
respectively. It is obvious in the latter case that 
the ${\rm Re}\delta_B$ contribution to ${\cal A}(t)$
is significant (even dominant for large $\Gamma t$)
and a separate bound on ${\rm Im}\delta_B$ cannot be obtained.

\vspace{0.5cm}

Next let us simplify the result in Eq. (8) by taking
$\cos\theta =0$ or $\sin\theta =1$, i.e., no $CPT$ violation
in $B^0$-$\bar{B}^0$ mixing involved. In this case 
we obtain
\begin{eqnarray}
{\rm Re}(\Omega + \bar{\Omega}) & = &
2 \sinh ({\rm Im}\phi) \left [
{\rm Im}Z_{ll'} \sin ({\rm Re}\phi)
- {\rm Re}Z_{ll'}
\cos ({\rm Re}\phi) \right ] \; , \nonumber \\
{\rm Im}(\Omega + \bar{\Omega}) & = &
2 \cosh ({\rm Im}\phi) \left [
{\rm Im}Z_{ll'} \cos ({\rm Re}\phi)
+ {\rm Re}Z_{ll'} \sin ({\rm Re}\phi) \right ] \; ,
%		(14)
\end{eqnarray}
where $Z_{ll'} \equiv \sigma^{~}_l + \sigma^{~}_{l'}$
with $|Z_{ll'}|\ll 1$. Note that
${\rm Im}\phi$ describes the $CP$-violating effect
induced by $B^0$-$\bar{B}^0$ mixing and can in principle 
be measured (or constrained) from the same-sign dilepton 
asymmetry of neutral-$B$ decays (see, e.g.,
Refs. \cite{Sanda,Xing97}).
The magnitude of ${\rm Im}\phi$ is expected to be
of $O(10^{-3})$ within the standard model, but it
might be enhanced to $O(10^{-2})$ if there exists the 
$CP$-violating new physics in $B^0$-$\bar{B}^0$ mixing
\cite{Xing98}. Anyway the ${\rm Re}(\Omega + \bar{\Omega})$ 
term is doubly suppressed by ${\rm Im}\phi$
and $|Z_{ll'}|$, therefore it can be
neglected in the rate difference between
${\cal R}(t)$ and $\bar{\cal R}(t)$. The asymmetry
${\cal A}' (t)$, defined as the ratio of ${\cal R}(t)-
\bar{\cal R}(t)$ to ${\cal R}(t)+\bar{\cal R}(t)$,
is then governed by the ${\rm Im}(\Omega + \bar{\Omega})$
term with $\cosh ({\rm Im}\phi)
\approx 1$ as follows:
\begin{equation}
{\cal A}' (t) \; =\; 2 \frac{\left [{\rm Im}Z_{ll'}
\cos ({\rm Re}\phi) +
{\rm Re}Z_{ll'} \sin ({\rm Re}\phi)
\right ] \sin (x\Gamma t)}{\cosh (y\Gamma t) + \cos (x\Gamma t)} \; \; .
%		(15)
\end{equation}
This instructive result implies
that a signal of $\Delta B =-\Delta Q$ transitions
can emerge from the opposite-sign dilepton asymmetry
of neutral-$B$ decays, if $CPT$ violation is absent or
negligibly small. Whether the effect of $\Delta B =\Delta Q$
violation is significant or not beyond the standard model,
however, remains an open question.

Comparing Eq. (15) with Eq. (12), one can see that
the neglect of the $\sinh(y\Gamma t)$ term in
${\cal A}' (t)$ is much safer than that in ${\cal A}(t)$.
In general, the ${\rm Re}(\Omega + \bar{\Omega})$ 
and ${\rm Im}(\Omega + \bar{\Omega})$ terms 
may consist of comparable effects
from $CPT$ violation and $\Delta B =-\Delta Q$ transitions.
Hence a constraint on ${\rm Im}(\cos\theta)$
or ${\rm Im}\delta_B$, due to the presence of
$\Delta B =-\Delta Q$ contamination, becomes impossible
in the opposite-sign dilepton events.

\vspace{0.5cm}

Keeping with the current experimental 
interest in tests of discrete symmetries and conservation
laws at the upcoming $B$ factories, 
we have made a new analysis of the opposite-sign
dilepton decays of neutral-$B$ mesons.
It is shown that the effects of 
$\Delta B = -\Delta Q$ transitions and $CPT$ violation
have the same time-dependent behavior in
such dilepton events, hence they are in general 
indistinguishable from each other.
To separate one kind of new physics from the other
will require
some delicate measurements of the opposite-sign
dilepton events, the same-sign dilepton events, 
and other relevant nonleptonic decay channels.
We remark that an experimental bound on 
the $CPT$-violating parameter ${\rm Im}\delta_B$ relies
on the assumptions $y\approx 0$ and $|{\rm Re}\delta_B|
\leq |{\rm Im}\delta_B|$ as well as the validity of the
$\Delta B =\Delta Q$ rule. For this reason
the value obtained in Ref. \cite{OPAL} can only shed limited light
on $CPT$ invariance or its possible violation in
the $B^0$-$\bar{B}^0$ mixing system. However, a new
analysis of the old data is likely to give a constraint
on the imaginary part of $(\Omega + \bar{\Omega})$,
provided its real part is not significant.

We hope that a useful test of $CPT$ and $\Delta B =\Delta Q$
conservation laws can finally be realized at the
asymmetric $B$-meson factories. The prospect of such
ambitious experiments, as discussed in Ref. \cite{Sakai98},
is not dim. And the relevant theoretical motivation seems
encouraging too \cite{CPT}.
On the phenomenological side, 
further and more general analyses
of $CP$-violating, $CPT$-violating and $\Delta B =-
\Delta Q$ effects on all types of neutral $B$-meson decays  
are therefore desirable.

\vspace{0.5cm}

The author likes to thank M. Jimack for his helpful comments
on this work. He is also grateful to V.A. Kosteleck$\rm\acute{y}$,
L.B. Okun, Y. Sakai and S.Y. Tsai for useful discussions.

\newpage

\begin{figure}
% GNUPLOT: LaTeX picture
\setlength{\unitlength}{0.240900pt}
\ifx\plotpoint\undefined\newsavebox{\plotpoint}\fi
\sbox{\plotpoint}{\rule[-0.175pt]{0.350pt}{0.350pt}}%

\vspace{0.4cm}
\caption{Illustrative plot for the time distribution of ${\cal A}(t)$
with $r \equiv {\rm Re}\delta_B/{\rm Im}\delta_B = \pm 1$ or $\pm 100$, 
where $x=0.7$ and $y=0.01$ have typically been taken. }
\end{figure}

\end{document}